\documentclass[nofootinbib,showpacs,floatfix,prc]{revtex4}
\usepackage{amsfonts}
\usepackage{graphicx}
\usepackage{amsbsy}
\usepackage{longtable}
\newcommand{\be}{\begin{equation}}
\newcommand{\ee}[1]{\label{#1} \end{equation}}
\newcommand{\ba}{\begin{eqnarray}}
\newcommand{\ea}[1]{\label{#1} \end{eqnarray}}
\newcommand{\nl}{\nonumber \\}
\newcommand{\re}[1]{(\ref{#1})}
\newcommand{\ep}{\epsilon}

\newcommand{\pd}[2]{\frac{\partial #1}{\partial #2}}

\begin{document}

\title{A thermodynamic approach to the relaxation of viscosity and thermal conductivity}

\author{T. S. B\'ir\'o$^{1}$, E. Moln\'ar$^{2}$ and P. V\'an$^{1}$ }
\affiliation{
$^1$ MTA-KFKI, Research Institute of Particle and Nuclear Physics,
H-1525 Budapest 114, P.O.Box 49, Hungary\\
$^2$ Frankfurt Institute for Advanced Studies, J. W. Goethe University, Max-von-Laue-Str. 1,
D-60438 Frankfurt am Main, Germany
}


\begin{abstract}
A novel higher order theory of relaxation of heat and viscosity is proposed based on corrections to the traditional treatment of the relativistic energy density. In the framework of generalized Bjorken scaling solution to accelerating longitudinal flow we point out that the energy flux can be consequently set to zero in the stationary case, independently of the choice of a specific local rest frame, like the Landau-Lifshitz or Eckart one. We investigate and compare several cooling and re-heating scenarios for the Quark Gluon Plasma (QGP) within this approach.
\end{abstract}

\maketitle

\date{25.04.2008}

\section{Introduction}

The fluid dynamical description of the evolution of strongly interacting matter created in heavy-ion
collisions, initially proposed and applied to describe p+p collision at low energies, was
pioneered by Landau \cite{Landau_1}. Ever since then, it has been successfully used to
model different colliding heavy ions at a wide range of
energies. Nowadays, one of the most intriguing and important experimental discoveries at
the Relativistic Heavy Ion Collider (RHIC) in Brookhaven, US, the measurement of collective flow in
non-central Au+Au collisions, demonstrates the predicting power of the fluid dynamical
approach.

Experimental evidence in single-particle transverse momentum distributions,
like radial flow, and in the coefficients of the asymmetric azimuthal distribution
around the beam axis, the directed transverse flow $v_1$, the elliptic
flow $v_2$ and the anti-flow $v_3$, shows that the predictions of perfect fluid dynamical models
assuming initial conditions from the Color Glass Condensate (CGC) \cite{Hirano_1},
overestimate certain data \cite{Hirano_2}.
In particular, the elliptic flow, $v_2$, surmised to be created in the early stage
of the collision signaling an early onset of thermalization, could only be reproduced
- using perfect fluid dynamical calculations with an initial condition  of Glauber type - with
thermalization time $\tau_0 < 1$ fm/c, up to transverse momenta $p_{\perp} \leq 1.5$ GeV \cite{Heinz-Hirano}.

Remarkably, the perfect fluid dynamical calculations with the CGC initial state using
a realistic description of the dissipative hadronic corona, could still not reproduce the
elliptic flow data \cite{Hirano_4}. This instigates that additional dissipation must
happen in the fluid dynamical stage: the matter created in high energy heavy ion
collisions can not be completely described by perfect fluid dynamics with zero
viscosities and without heat conduction.

Triggered by these developments there is an increasing interest in relativistic
dissipative fluid dynamics. There are several recent suggestions and modifications and
renewed discussions of the old enigmas of relativistic viscous fluids \cite{NewHydro,BaiAta07m,TsuKun08m}.
The most intensely investigated problems are the stability of the homogeneous equilibrium solutions and the
causal propagation of perturbations.
These issues are related to each other.

There are several investigations connecting causality and stability in dynamical systems described by hyperbolic
partial differential equations. According to these results the mathematical  properties of symmetric hyperbolic
equations ensure that the propagation speed of perturbations is finite. Due to certain additional restrictions
on the material properties, these characteristic speeds are less than
the speed of light. For the so called divergence type theories \cite{GerLin90a} it is straightforward
to determine the conditions of causality for the full set of nonlinear evolution equations.
Moreover, for these theories causality implies stability. In case of
Israel-Stewart fluids it has been proven that the symmetric hyperbolicity of the perturbation equations
is equivalent to the conditions of linear stability of the homogeneous equilibrium \cite{HisLin83a,HisLin88a}.
However, the original non-perturbed equations in these theories are
not known to be symmetric (let alone causal) for arbitrary fluid states.

On the other hand, stability also has certain implications on causality. If the homogeneous equilibrium is
asymptotically stable, then the causality region of the theory shall be reduced due to damping.
The causality region can be restricted by physical characteristic
speeds \cite{ParaCaus,BiroVan}. In this sense relativistic {\em parabolic} theories are viable,
provided one can prove the stability of the homogeneous equilibrium.
Several authors argue that hyperbolic extensions of the
Navier-Stokes-Fourier system would not have experimental consequences; the essential part of
dissipative relativistic hydrodynamics is the parabolic Navier-Stokes-Fourier core \cite{HypPara}.
According to these arguments the proof of the stability is a most fundamental issue in dissipative theories.

As it is well known, the so called first order theories are unstable \cite{1OrdInst}.
The  stability of the homogeneous equilibrium in several recent, second order
theories either was not investigated \cite{NewHydro,BaiAta07m}, or the
obtained stability conditions seem to be too restrictive \cite{TsuKun08m}.
It is important to note, that the stability conditions in the Israel-Stewart theory are complicated,
and cannot be conceived in a natural way, see for example Eq. (70) in Ref. \cite{HisLin83a}.
This circumstance is in strong contrast to the nonrelativistic case, where the thermodynamic (equivalently hydrodynamic)
stability and the positivity of the viscosities and the heat conduction
coefficient ensure the linear stability of the homogeneous equilibrium without any further elaboration.

We have recently analyzed \cite{BiroVan} the physical reasons of instabilities in the first order theory of Eckart \cite{Eckart}.
Our conclusion was that stability of the homogeneous equilibrium can be restored
independently of the chosen frame (Eckart or Landau-Lifshitz), by exploring the physical
difference between momentum density and energy flux in the local rest frame.
We have given a minimal, stable extension of the first
order Eckart theory by correcting the traditional treatment of the energy density.
Unlike in previous relativistic theories of dissipative fluids, where the local rest frame energy is considered as
the internal energy, we have suggested to apply the absolute value of the local rest frame energy
vector, $\tilde e = \sqrt{u_a T^a_b T^b_cu^c}$.
We proved that in this case the positivity of the coefficients in the classical linear response theory
(heat conduction coefficient and viscosities) and the conditions of thermodynamic stability are sufficient
to avoid generic instabilities. In this theory no further conditions are needed, in full analogy to the
non-relativistic Navier-Stokes-Fourier equations.

In this paper our aim is to extend this approach to a set of  hyperbolic equations.
Our starting point is the fiducial equation of state between the entropy density and the energy density.
In this relation we use Lorentz scalar combinations of the energy-momentum tensor at
zero pressure and the flow four-velocity. The analysis will be carried out
without specifying the flow frame.  We derive the relaxation equations of heat flux and viscous
pressures from the corresponding entropy production. Then we investigate  a certain generalization of
the Bjorken flow and show that the stationary solution implies vanishing heat flux.
Finally we solve the subsequent equations and investigate the correspondence of expansion,
cooling and re-heating in our approach.

\section{Thermodynamics}

In this section we derive the equations of viscous fluid dynamics using the convention
where the upper indices denote contravariant while the lower indices covariant four vectors.
The metric tensor is given as $g_{ij}={\rm diag}(1,-1,-1,-1)$ and all indexes $i,j,k, ..$ run over $0,1,2,3$.
We use natural units, $\hbar=k=c=1$, except the final section.

The projection of the energy-momentum tensor, $E^j=u_iT^{ij}$ can be interpreted as the energy flux in the local
comoving frame, the scalar projection, $e =u_iT^{ij}u_j=E^ju_j$, comprises the local, relativistic energy density.
We have studied the modified equation of state, $ s(\tilde e,n)$, with
${\tilde e}^2=E_iE^i$ in Ref. \cite{BiroVan} and found that it leads to a
stabilization of the known generic instabilities.
The change of variables of the entropy density was supported by arguments from modern non-equilibrium
thermodynamics based on the Liu procedure applied to first order weakly nonlocal state spaces
\cite{Van07a}. The original Eckart theory considers both time-space non diagonal  components
of the energy-momentum tensor, the energy flux and the momentum density, as dissipative contributions.
We have argued that using ${\tilde e}$  as internal energy density, results in a distinction in their physical role,
and restores the stability of the homogeneous solutions.

Now we extend this approach by utilizing other Lorentz scalars in
combinations transverse to $u^i$. With the energy like Lorentz scalars, ${e}^2$,
$E_iE^i$ and $T_{ij}T^{ij}$, we construct a fiducial energy density expression containing all the dissipative and heat conducting terms and derive relaxation equations for them. The energy-momentum tensor can be generally decomposed \cite{Eckart} into reversible and irreversible parts. With the help of the transverse projector, $\Delta^{ij}=g^{ij}-u^iu^j$, we consider
\be
 T^{ij} = eu^iu^j + q^iu^j + u^iq^j - (p+\Pi) \Delta^{ij} + \pi^{ij},
\ee{EMOM}
where we define the transverse energy flow four vector (or heat flow in case
one uses Eckart's definition), as $q^i = \Delta^{i}_{\, k} T^{kj} u_{j} $,
the hydrostatic pressure, $p$, the bulk viscous pressure $\Pi$,
where $(p+\Pi) = -\frac{1}{3}\Delta_{ij} T^{ij}$, and the shear stress tensor, $\pi^{ij} = T^{<ij>} - (p + \Pi)\Delta^{ij}$.
The $T^{<ij>} = \frac{1}{2}(T^{kn} + T^{nk})\Delta_k^i\Delta_n^j - \frac{1}{3} \Delta^{ij}\Delta_{kn}T^{kn}$
notation stands for a particular, symmetrized and traceless combination of indices reflecting the same property of the shear stress tensor, $\pi_{ij}$.

The energy flux four vector is hence given by
\be
 E^i =  eu^i + q^i,
\ee{ENER}
while its Lorentz invariant square length becomes
\be
 E_iE^i =  e^2 + q_iq^i.
\ee{E2}
We note that since $q_iu^i=0$ and $u^i$ is a timelike vector, $q^i$ is spacelike,
so $q_iq^i \le 0$. The Lorentz scalar square of the energy-momentum tensor reads as
\be
 T_{ij}T^{ij} = {e}^2 + 2q_iq^i + 3(p+\Pi)^2 + \pi_{ij}\pi^{ij},
\ee{TENSOR2}
where the number $3=\Delta^{i}_{i}$ reflects the space dimensions. Our present ansatz generalizing
the local invariant energy density including irreversible processes is given by:
\be
 L = E_iE^i + (E_iE^i-T_{ij}T^{ij}) + (E_iE^i-{e}^2).
\ee{ANSATZ} Here the first two terms, $2E_i E^i - T_{ij} T^{ij} =  e^2 - 3(p+\Pi)^2
- \pi^{ij} \pi_{ij}$, contain the square of the energy density and the dissipative
contributions due to the hydrostatic pressure plus bulk pressure and shear stress. The
last term, $E_iE^i-{e}^2$, returns the square of the absolute value of the heat
flow. Substituting these values our fiducial scalar entering the equation of state is
given by the square root of
\be
L =e^2 + q_iq^i - 3(p+\Pi)^2 - \pi_{ij}\pi^{ij}.
\ee{OURENER}
We interpret this construction as the following physical picture: The effective energy
density ${\ep}=\sqrt{L(p=0)}$ is equal to the familiar one, $e$, in the absence of
dissipation. Dissipation decreases $\ep$ compared to $e$, and since the
entropy is a monotonic growing function of internal energy for systems with positive
absolute temperature, the entropy is maximal at no dissipation. This construction is akin
to the M\"uller and Israel-Stewart approach in its spirit \cite{RelExtOri}.
The $p=0$ version of eq. (\ref{OURENER}) separates the static from the dissipative parts of
the pressure, according to the basic presumption of these theories.
In our approach the equation of state is given by a particular function,
\be
 \hat s(\ep,n) \equiv \hat s\left(\sqrt{{e}^2 + q_iq^i - 3\Pi^2 - \pi_{ij}\pi^{ij}},n \right)
 = s\left({e},q^i,\Pi,\pi^{ij},n \right).
\ee{EOS}
For thermodynamical systems without heat conduction and dissipative effects, $q_i=0$, $\Pi=0$ and $\pi_{ij}=0$,
the traditional  equilibrium relation emerges, $\hat s(\ep,n)|_{q^i=0,\Pi=0,\pi^{ij}=0} \equiv \hat{s}(e, n) = s(e,0^{i},0,0^{ij},n)$.

Moreover, $\hat s(\ep,n)$ achieves maximum at this point; so it is ensured that near
equilibrium the dissipative currents relax.
An expansion of $\ep$ for small dissipative currents to leading order leads to an
Israel-Stewart type of ansatz, however with fixed coefficients of the quadratic terms,
\be
 \hat s(\ep,n) \approx \hat s(e,n) + \frac{1}{2 e}
\left( q_i q^i - 3\Pi^2 - \pi_{ij}\pi^{ij}\right)\frac{\partial \hat s(e,n)}{\partial e} \, +...
\ee{ISRAEL}
Here identifying the inverse equilibrium temperature via $1/T_{eq}=\partial \hat s(e,n)/\partial e\ $,
the Israel-Stewart coefficients are given by $\beta_0=3/e$ for the bulk viscosity,
$\beta_1=1/e$ for the shear viscosity and $\beta_2=1/e$ for the heat flow.
This approach differs from the one obtained in the framework of kinetic theory,
nevertheless for an ideal relativistic gas ($e=3p$) the coefficients are all inversely
proportional to the pressure, and our result comes close to some of the coefficients in Ref. \cite{Israel_Stewart}.
Furthermore our formula can be applied for matter containing massless particles, while some
results calculated from relativistic kinetic theory in Ref. \cite{Israel_Stewart} diverges for $m=0$,
as well as for vanishing pressure.
Here we note that one might introduce different scalar functions in front of all newly introduced
scalar terms when constructing the effective energy density, for example in order to match the thermodynamical coefficients introduced by Israel and Stewart. In the present paper we do not introduce such coefficients in order to keep the investigations simple and transparent.

In the following we study the class of relativistic equation of states involved in eq. (\ref{EOS}) without, surpassing this way the usual second order fluid dynamics of Israel-Stewart. We note that it is useful to introduce a common notation for all dissipative modifications. We consider $\ep^2=e^2-D^2$, where $D^2 = -q_i q^i + 3\Pi^2 + \pi_{ij}\pi^{ij}$ from which the approximations, $\ep \approx e - D^2/(2e)$
and $ \hat s(\ep,n) \approx \hat s(e,n) - D^2/(2eT) $ holds.

The Gibbs relation can be obtained by inspecting the total differential of the entropy
density (\ref{EOS}). We obtain two sets of intensive variables differentiating the indirect function,
$\hat s(\ep(e,q_i,\Pi,\pi_{ij}),n)$.
The differential with respect to $\ep$ and $n$,
\be
 d\hat s(\ep,n) \equiv \pd{\hat s}{\ep}d\ep + \pd{\hat s}{n} dn = \frac{1}{\theta} d\ep - \frac{\hat\mu}{\theta} dn \, ,
\ee{no1}
defines the effective temperature $\theta$ and effective chemical potential $\hat \mu$ by simple expressions.
The introduction of these new intensive variables naturally mimics the standard thermodynamic relations in case of equilibrium, 
the interpretation of these quantities becomes clear in the equilibrium limit.
  
On the other hand, since $\ep$, depends on $e, q^i, \Pi$ and $\pi^{ij}$,
\begin{eqnarray}
 ds(e,q^i,\Pi,\pi^{ij},n) &\equiv&
 \pd{\hat s}{\ep}
 \left(\frac{e}{\ep}de +\frac{q_i}{\ep}dq^i-\frac{3\Pi}{\ep}d\Pi
    -\frac{\pi_{ij}}{\ep}d\pi^{ij} \right)
    + \pd{s}{n} dn  \label{DIFFENT}\\
 &=&
\frac{1}{T} d e +
        \frac{q_i}{e T}dq^i-
        \frac{3\Pi}{e T}d\Pi -
        \frac{\pi_{ij}}{e T}d\pi^{ij}-
        \frac{\mu}{T} dn \, ,\nonumber
\end{eqnarray}

\noindent enables us to introduce intensive variables associated to $de$, $dq^i$, $d\Pi$, $d \pi^{ij}$ and $dn$, respectively.
As we will see, these intensive quantities appear in the  diffusion, pressure  and heat conduction relaxation equations, driving the  thermodynamic system toward equilibrium. Therefore $T$ is called \textit{equilibrating temperature} distinct from the \textit{effective temperature} $\theta$ appearing in the state functions. Using the above relations we can easily establish the following relations between those quantities
 \be
 e T = \ep\theta, \qquad \text{and} \qquad e\mu = \ep\hat\mu \, ,
 \ee{relT}
where the functions $\theta$, $T$ and $\hat\mu$, $\mu$ are respectively equal and reduce to the equilibrium
temperature and chemical potential, $T_{eq} = T = \theta$ and $\mu_{eq} = \mu = \hat\mu$, when the energy
flux and the viscous pressure vanish $(q^i=0, \Pi=0, \pi^{ij}=0)$.
The equilibrium entropy density corresponds to the previously introduced $\hat s(e,n)=s_{eq}(e,n)$, because $de = T_{eq} ds_{eq}+\mu_{eq}dn $.

\section{Thermodynamic pressure}

Since the derivative with respect to proper time in a frame comoving with the local flow
is given by $d/d\tau=u_i\partial^i$, the above total derivative multiplied with the
temperature can be expressed as
 \be
 T\frac{ds}{d\tau} = T\partial_i(su^i) - Ts\partial_iu^i.
\ee{TAUDERIV}
This leads to the following entropy balance equation
 \be
 T\partial_i(s u^i) = \partial_i(eu^i)+\frac{q_j}{e}\partial_i(q^ju^i)
 - \frac{3\Pi}{e}\partial_i(\Pi u^i) - \frac{\pi_{jk}}{e}\partial_i(\pi^{jk}u^i)
 - \mu \partial_i(nu^i) + \hat p  \partial_i u^i.
\ee{HEAT}
Here each term expresses a contribution to the entropy production as a product of
an intensive parameter and the divergence of an extensive one. In particular the
last term in eq. (\ref{HEAT}) contains the mechanical work on a changing volume
(considering $dV/d\tau=V\partial_i u^i$). The coefficient ${\hat p}$ turns out to be
\be
\hat p = T\hat s(\ep,n) + \mu n - \frac{\ep^2}{e}.
\ee{PTILDE}
An equivalent way reminding the familiar thermodynamic relation is given as
\be
 \frac{\theta}{T} \hat p \: = \: \theta \hat s(\ep,n) + \hat\mu n - \ep.
\ee{ANOTHER_P_TILDE
}

Using the comprised notation $D^2$ for the sum of dissipative modifications in the
energy density, $\ep^2 = e^2 - D^2$, while expanding the entropy density around the traditional
formula, $\hat s(\ep,n) = s_{eq}- D^2/(2eT_{eq})$, an approximate formula arises for the pressure
\be
 \hat p \approx T_{eq} s_{eq} + \mu_{eq} n -e + \frac{D^2}{2e} \, .
\ee{PRESSEXPAND}
On the other hand for the - till this point unspecified -
parameter $p$  in the energy-momentum tensor in the dissipative case we suggest to use the standard expression,
 \be
 p = T\hat s+\mu n - e \, .
\ee{SIMPLEP}
With this definition in general it follows that,
$\hat p = p + D^2/e \approx  p_{eq} + D^2/(2e)$, with $p_{eq} = T_{eq}s_{eq}+\mu_{eq} n-e$ being the isotropic equilibrium pressure in the absence of dissipation.

At this point it is a delicate question what part of the total
pressure will be assigned to dissipative and what to non-dissipative
effects; the thermodynamic interpretation of the parameter $p$
occurring in the energy-momentum tensor has to be determined. While in
the absence of dissipation $p$ coincides with the isotropic,
equilibrium hydrostatic pressure, $p_{eq}$, from the analysis of the
mechanical work done on the expanding volume the quantity $\hat p$
is relevant. We motivate our choice by the following
brief analysis of the example of pure radiation.

Our starting point is the following equation of state for an ideal gas of massless particles (radiation):
\be
 \hat s(\ep) = s(e,q^i,\Pi,\pi^{ij})=a\ep^{3/4},
 \ee{RADEOS}
\noindent where $\sigma=(3 a/4)^4$ is the Stefan-Boltzmann
constant. The inverse temperatures in this case are given by
 \be
 \frac{1}{T} = \pd{s}{e} = \frac{e}{\ep} \: \frac{3}{4} \: a \: \ep^{-1/4}, \qquad
 \frac{1}{\theta} = \pd{\hat s}{\ep}=\frac{3}{4} \: a \: \ep^{-1/4} \, .
\ee{RADINVTEMP}
 This means that the effective energy density is, $\ep=\sigma \theta^4$, while the pressure from eq.(\ref{PTILDE}) becomes
\be
 \hat{p} \equiv \frac{1}{3} \frac{\ep^2}{e}
= \frac{1}{3} \: \ep \: \frac{T}{\theta}.
\ee{RADPRESS}
From this at no dissipation the familiar equation of state $p_{eq}=e/3$ arises. From the energy balance one knows, that the cooling of an expanding system is driven by the quantity $h=e+p$.
For the pure radiation this becomes
\be
 h= e + p = T\hat s = \frac{4}{3} \frac{\ep^2}{e} = 4 \hat{p}.
\ee{RADTHALP}
From this study we conclude that the interpretation of $\ep$ as internal energy must be developed consequently.
In our understanding the thermodynamic pressure is given by $\hat p$ in eq. \re{PTILDE}.

\section{Entropy production and linear response}

In order to proceed further by determining the entropy  production we need the energy-momentum balance,
\be
  \partial_j T^{ij} = 0.
\ee{CONSERVEDENERG}
This way using eq. \re{EMOM}, the energy balance can be expressed as
\be
 u_i\partial_jT^{ij}=\partial_i(eu^i) + \partial_i q^i +u_i\dot{q}^i + (p+\Pi)\partial_iu^i - \pi^{ij} \partial_i u_j =0.
\ee{ENERGFLUX}
Here the over-dot denotes the proper time  derivative, $\dot{x}=dx/d\tau=u_i\partial^i x$.
Using the fact that $q_iu^i=0$ we replace
the term $u_i\dot{q}^i=-q_i\dot{u}^i$.  The Euler-equation is given
 \be
      \Delta^i_{\;\;k}\partial_j T^{kj} =
        e\dot{u}^i  +
        {q}^i \partial_j u^j +
        q^j \partial_j u^i +
        \Delta^i_{\;\;k} \dot{q}^k
        - \Delta^i_{\;\;k}\partial_j P^{kj} = 0,
 \ee{ibal}
\noindent where $P^{ij}=(p+\Pi)\Delta^{ij}+\pi^{ij}$.

Substituting the energy balance into the
entropy balance (\ref{HEAT}) we obtain
 \be
T\partial_i(su^i)= - \partial_i q^i + q_i\dot{u}^i - \Pi \partial_i u^i +
\pi^{ij} \partial_i u_j -\mu\partial_i(nu^i) + \frac{q_i}{e}\dot{q}^i - \frac{3\Pi}{e}\dot{\Pi} - \frac{\pi_{ij}}{e}\dot{\pi}^{ij}.
\ee{sbal}

The particle four current is defined generally as $N^i=n u^i + \nu^i$, where $\nu^i$ is the particle flux,
which is spacelike in the local rest frame.
Associating the chemical potential to a conserved particle, $\partial_i N^i=\partial_i (nu^i)+\partial_i\nu^i=0$, and it allows to use
$\partial_i(nu^i)=-\partial_i\nu^i$ in the entropy balance.
Therefore, substituting  eq. \re{ENERGFLUX} into eq. \re{sbal} we get
\begin{eqnarray}
 T \partial_i S^i &=&
        T\left(\partial_i(su^i)+\partial_i\left(\frac{q^i-\mu\nu^i }{T}\right)\right) = \frac{q_i}{T}\left(T\dot{u}^i + \frac{T}{e}\dot{q}^i + \partial^i T \right)-
\nonumber\\
 && \Pi \left(\partial_i u^i + \frac{3}{e} \dot{\Pi}\right)
 + \pi_{ij} \left( \partial^i u^j - \frac{1}{e} \dot{\pi}^{ij}\right)
 - T \nu_i \partial^i\left(\frac{\mu}{T} \right) \geq 0 \, ,
\label{ENTROFLUX}
\end{eqnarray}
where according to the Second Law of thermodynamics the entropy production is non-negative.
We can see, that convenient definition of the entropy flux is $j^i=(q^i-\mu\nu^i)/T$ as done by
Eckart in analogy to the nonrelativistic case.
Hence the entropy four current is defined as
\begin{eqnarray}
S^i &=& \hat s(\ep,n) u^i + \frac{q^i}{T} - \frac{\mu \nu^i}{T} \nl
&\approx & s(e,n) u^{i} + \frac{q^i}{T} - \frac{\mu \nu^i}{T}
+ \left(q_j q^j - 3\Pi^2 - \pi_{jk}\pi^{jk} \right)\frac{u^{i}}{2eT}.
\ea{FOURCURR}

Here the first term in the last line denotes the entropy of a perfect fluid carried by the flow,
the second term denotes the entropy flux due to heat and particle flux.
In case one uses Eckart's definition for the flow field, $\nu^i = 0$, the energy flux $q^i$ equals to the heat flow.
On the other hand using the definition of Landau and Lifshitz for the flow, $q^i = 0$, the energy flux vanishes in
the local rest frame and heat flux is defined by $I^i=- (e+p)/n \nu^i$.
In a baryon free matter one obviously has $\nu^i = 0$, therefore there is no temperature equilibration by heat conduction.

The construction involved in \re{FOURCURR} truncated terms up to linear order in dissipative quantities corresponds
to Eckart's {\em first order theory} of relativistic dissipative fluid dynamics.
The introduction of quadratic terms are generally referred to as
{\em second order theories} of dissipative fluid dynamics \cite{HisLin88a,Muronga_old}.
Our approach (cf. first line in eq. \re{FOURCURR}) contains an infinite series of higher order terms.

A complete set of second order terms in the
entropy four current of relativistic fluids was proposed by Israel and Stewart \cite{Israel_Stewart}, with coefficients
$\beta_0, \beta_1$ and $\beta_2$, for the quadratic terms in $\Pi$, $q^i$ and $\pi^{ij}$ respectively.
They introduced also $\alpha_0$ and $\alpha_1$ coefficients in front of viscous-heat flux coupling
terms like $\Pi q^{i}$ and $\pi^{ij} q_{j}$.
Since, our main ansatz is based on Lorentz scalar second order quantities involved in the dissipative
relativistic energy density, it is clear that such vector terms  do not appear in our present approach.

The M\"uller, Israel and Stewart method modifies and generalizes explicitly the
entropy four current of Eckart, through which the definition
of the local rest frame entropy density is extended.
They give the most general isotropic and second order expression in a pure mathematical way.
Our method is based on constructing local corrections to
the energy density, and then define the entropy
four current along the lines as done  by Eckart. However, expanding the corrections to
the equilibrium entropy density (\ref{FOURCURR}) results in a formula resembling the
definition of Israel and Stewart.

The nonnegativity of the entropy production in \re{ENTROFLUX} can be ensured if each term is separately non-negative.
Therefore in case of isotropic materials the coefficients of the energy flux
$q^i$ (orthogonal to $u^i$, i.e. the heat conduction), the bulk and shear viscosity
terms multiplying $\Pi$ and $\pi_{ij}$ in the above expression and the chemical diffusion
contribution are treated as being proportional to the corresponding coefficient in the
linear response approximation. This leads to the following equations:
\begin{eqnarray}
 q^i &=& - \lambda \left(T \Delta^{ij}\dot{u}_j
          + \frac{T}{e}\Delta^{ij}\dot{q}_j + \nabla^i T  \right), \label{Four}\\  
 \Pi &=& -\zeta \left(\partial_i u^i + \frac{3}{e} \dot{\Pi} \right), \label{Bvisc}\\
 \pi_{ij} &=& 2\eta \left(\partial_{<i} u_{j>} - \frac{1}{e} \dot{\pi}_{ij} \right), \label{Svisc}\\
 \nu^i &=& - \sigma \partial^i \frac{\mu}{T} ,
\ea{LINEAR}
where the short hand notation $\nabla^i=\Delta^{ij}\partial_j$ has been introduced.
The heat conductivity $\lambda$, the bulk viscosity $\zeta$ and the shear viscosity
$\eta$ are non-negative transport coefficients.
The diffusion term (last line in eq.(\ref{LINEAR})) and the heat conduction term
(first line) may in general show cross-couplings (cf. Soret effect \cite{GroMaz62b}).
With the above simple construction, we arrive at the following formula for the entropy production,
\be
 \partial_i S^i =
       \frac{\Pi^2}{\zeta T} - \frac{q_i q^i}{\lambda T^2} +
       \frac{\pi_{ij} \pi^{ij}}{2\eta T} - \frac{\nu_i\nu^i}{\sigma T}\geq 0 . 
\ee{ENTROPROD}

In the following discussions  we concentrate on the heat conduction and viscosity evolution and
neglect particle diffusion phenomena. Based on the above considerations we obtain the following 
evolution equations for the viscosity terms:
\ba
 \frac{1}{e}\Delta^{ij}\dot{q}_j + \frac{1}{T}\nabla^i T + \Delta^{ij}\dot{u}_j
                + \frac{1}{\lambda T} q^i &=& 0, \nl
 \frac{3}{e}\dot{\Pi} + \partial_i u^i + \frac{1}{\zeta} \Pi &=& 0, \nl
 \frac{1}{e}\dot{\pi}_{ij} - \partial_{<i} u_{j>} + \frac{1}{2\eta} \pi_{ij} &=& 0.
\ea{RELAX}
The corresponding relaxation times are hence given by
$\tau_q = \lambda T \beta_1 = \lambda T/e$ for the heat conduction,
by $\tau_{\Pi}= \beta_0 \zeta = 3\zeta/e$ for the bulk viscosity and
by $\tau_{\pi} = 2\eta \beta_2 = 2\eta/e$ for the shear viscosity.
These relaxation times are perfectly finite even for massless matter.

The above relaxation equations closely resemble the  truncated Israel-Stewart form.
Such forms of the relaxation equations are extensively utilized in quark matter
research \cite{BaiAta07m,BRW,Muronga_new}.

The main difference to our approach lies  in terms which contain the space-time derivative
of the thermodynamic coefficients multiplied by the flow.
As argued by Heinz et. al. \cite{Heinz}, these Israel-Stewart quantities are rather small, and one can
neglect them safely. In the case they were large, they become unbounded and
the system may destabilize. In our
approach such terms, containing derivatives of the Israel-Stewart coefficients, do not appear. However, these equations already capture the essential features of relaxation phenomena and we expect that it will result in a causal and stable theory.

These relaxation equations can be solved parallel to the energy flux
equation (\ref{ENERGFLUX}) and the Euler equation describing the
evolution of the flow $u^i$.

\section{Generalized Bjorken flow}


Let the basis of coordinates be given by the tetrad $e_a^i$, such
that $x^i=\tau e_0^i + r e_2^i$, and therefore \hbox{$dx^i=d\tau
e_0^i+\tau d\eta e_1^i + dr e_2^i+rd\phi e_3^i$.} The unit four
vectors satisfy the orthogonality relations \ba e_a^i e_b^j g_{ij}
&=& g_{ab}, \nl e_a^i e_b^j g^{ab} &=& g^{ij} \ea{TETRAD} with the
Minkowski metric tensor $g_{ij}={\rm diag}(1,-1,-1,-1)$. The flow
velocity field and an orthogonal spacelike field are given by \ba
 u^i &=& \gamma \left(e_0^i + ve_1^i \right), \nl
 N^i &=& \gamma \left(ve_0^i+e_1^i \right),
\ea{FIELDS} where $\gamma=1/\sqrt{1-v^2}$, $u_iu^i=+1$, $N_iN^i=-1$
and $u^iu^j-N^iN^j=e_0^ie_0^j-e_1^ie_1^j$ so $u^i$ and $N^i$ span
the same two-dimensional subspace of the spacetime as $e_0^i$ and
$e_1^i$. It is therefore natural to consider derivatives in the
direction of these vectors, we call them as 'dot' and 'grad': \ba
 {\rm dot} = u_i\partial^i &=& \gamma \left(\pd{}{\tau}+\frac{v}{\tau}\pd{}{\eta} \right), \nl
 {\rm grad} = N_i\partial^i &=& \gamma \left(v \pd{}{\tau}+\frac{1}{\tau}\pd{}{\eta} \right).
\ea{DOTGRAD} Four-divergences of the flow and its orthogonal are
given as \ba
 A &=& \partial_i N^i = \frac{\gamma v}{\tau} + \gamma^2 \, {\rm dot}(v), \nl
 B &=& \partial_i u^i = \frac{\gamma }{\tau} + \gamma^2 \, {\rm grad}(v).
\ea{FOURDIV} All partial derivatives can be expressed by these
quantities: \ba
 \partial^i u^j &=& \left( Au^i - BN^i\right) N^j, \nl
 \partial^i N^j &=& \left( Au^i - BN^i\right) u^j.
\ea{PARC}
The transverse projection tensor in our basis is given by \be
 \Delta^{ij} =g^{ij}-u^iu^j= -\left(N^iN^j+e_2^ie_2^j+e_3^ie_3^j \right).
\ee{TRV} It has the properties $u_k\Delta^{ki}=0$ and
$N_k\Delta^{ki}=N^i$. This helps to obtain the traceless symmetric
part of the derivative tensor of the velocity field, important to
entangle the shear pressure term: \be
     \partial^{\langle i}u^{j \rangle} \: = \:
 B \left( -N^iN^j - \frac{1}{3}\Delta^{ij}\right) \: = \:
 B\left(-\frac{2}{3}N^iN^j+\frac{1}{3}e_2^ie_2^j+\frac{1}{3}e_3^ie_3^j \right).
\ee{SHEARDERIV} The shear pressure tensor $\pi^{ij}$ is proportional
to the expression in the bracket, we use a factor $3/2$ in the
definition \be
 \pi^{ij} = \pi(\tau,\eta) \left(-N^iN^j+\frac{1}{2}e_2^ie_2^j +\frac{1}{2}e_3^ie_3^j \right).
\ee{TENSORSHEAR} Now $\pi^{ij}\pi_{ij} = 3\pi^2/2$ follows. The usage of the $(u^i,N^i)$ reference frame allows
us to give the energy momentum tensor \re{EMOM} as follows: \be
 T^{ij} = eu^iu^j+q(N^iu^j+u^iN^j)+\alpha N^iN^j + \beta\left(e_2^ie_2^j+e_3^ie_3^j \right),
\ee{TANSATZ} with $\alpha=p+\Pi-\pi$ and $\beta=p+\Pi+\pi/2$. It
consists of projector terms, but the term proportional to $q$, which
is due to the $q^i=q(\tau,\eta)N^i$ form.

We consider the energy momentum conservation, including all
above terms in the $\partial_iT^{ij}=0$ equation. Its general form
is given by\be
 u_i {\rm dot}(T^{ij}) = N_i {\rm grad}(T^{ij}),
\ee{TFORM} so all equations describe a balance between 'dot' and
'grad' terms. In the case of the Bjorken flow these operations
coincide with the time and rapidity derivatives, but for a longitudinal accelerating
flow ansatz not. Together with the relaxation equations due to the
linear response assumption we obtain the following set of dynamical
equations corresponding to \re{ENERGFLUX}, \re{ibal}, \re{Four}, \re{Bvisc} and \re{Svisc} respectively:
\begin{eqnarray}
 {\rm dot}(e) + (e+\alpha)B +2qA+{\rm grad}(q) &=& 0, \nonumber\\
 {\rm dot}(q) + (e+\alpha)A +2qB+{\rm grad}(\alpha) &=& 0, \nonumber\\
 \frac{\lambda T}{e} {\rm dot}(q) + q + \lambda T A + \lambda {\rm grad}(T) &=& 0, \nonumber\\
 \frac{3 \zeta}{e} {\rm dot}(\Pi) + \Pi + \zeta B &=& 0, \nonumber\\
 \frac{2 \eta}{e} {\rm dot}(\pi) + \pi  - \frac{4\eta}{3} B &=& 0.\label{e5}
\end{eqnarray} Here the first equation describes the cooling due to
expansion, the second is the Euler equation describing the
acceleration of the flow due to pressure gradients, the third is the
Fourier heat conduction equation supplemented with a relaxation
term, while the fourth and fifth equations describe the relaxation of bulk and shear
viscosity.

It is interesting to consider that class of solutions when the
quantities under investigation depend only on the time variable,
$\tau$. Then denoting by $\dot{f}=df/d\tau$ for such functions we
arrive at \ba
 \dot{e} &=& -(e+\alpha)\tilde{B}-2q\tilde{A}-v\dot{q}, \nl
 \dot{q} &=& -(e+\alpha)\tilde{A}-2q\tilde{B}-v\dot{\alpha}, \nl
 \dot{q} &=& - \frac{e}{\lambda T\gamma}q -e\tilde{A} - \frac{ev}{T}\dot{T}, \nl
 \dot{\Pi} &=& - \frac{e}{3\zeta\gamma}\Pi - \frac{e}{3}\tilde{B}, \nl
 \dot{\pi} &=& - \frac{e}{2\eta\gamma}\pi +\frac{2e}{3}\tilde{B},
\ea{ONLYTAU} with \be
 \tilde{A} = \frac{v}{\tau} + \gamma^2 \dot{v}, \qquad {\rm and} \qquad
 \tilde{B} = \frac{1}{\tau} + \gamma^2 v \dot{v}.
\ee{TILDEAB} In this form the cooling and the Euler equation show a
quite symmetric role in the evolution, and furthermore it seems that
one would obtain two equations for $\dot{q}$. However, the Euler
equation has to be used to describe the change of the flow, so it
has to be regarded as an equation for $\dot{v}$ included in the
variables $\tilde{A}$ and $\tilde{B}$. The $v=0$ assumption reveals
that $\tilde{A}=\dot{v}$ and $\tilde{B}=1/\tau$, so indeed the roles
are not symmetric. In this case we obtain from the Euler equation
\be
 \dot{v} = - \frac{1}{e+\alpha} \left(\dot{q}+\frac{2q}{\tau} \right),
\ee{ZEROVEULER}
and then using the Fourier equation for eliminating
$\dot{q}$ from the above result we arrive at
 \be
 \dot{v} = \frac{q}{\alpha} \left( \frac{e}{\lambda\gamma}-\frac{2}{\tau}\right).
\ee{VDOTZEROV}
One concludes that only for $q=0$ can the $v=0$
Bjorken flow remain stationary. This on the other hand consequently
solves the Fourier equation. Hence we pointed out, that the energy flux, $q$,
can be consequently set to zero when considering the stationary Bjorken flow independent
of the Eckart or Landau-Lifshitz choice of the flow frame.

\newpage
\section{Expansion, cooling and re-heating in the scaling solution}

It is customary to investigate the relaxation of viscosity by utilizing
a stationary solution of the non-dissipative system, in particular for the
 quark gluon plasma evolution the Bjorken flow pattern \cite{Bjorken:1982qr}.
We have seen already that the only assumption consequent with a stationary Bjorken flow ($v=0$) is $q=0$.
In this case the acceleration of the flow and the heat current remain zero, due to
eqs. (\ref{ZEROVEULER}, \ref{VDOTZEROV}), while the energy density, the bulk and shear viscosity
relax in a coupled manner. 
Here, eqs. \re{ONLYTAU} simplifies to
\ba
  \dot{e} + \frac{e+p}{\tau} &=& \frac{\pi-\Pi}{\tau} \, , \nl
  \dot{\Pi} + \frac{e}{3\zeta}\Pi &=& - \frac{e}{3\tau} \, , \nl
  \dot{\pi} + \frac{e}{2\eta}\pi  &=& \frac{2e}{3\tau} \, .
\ea{EVOLQGP}


The corresponding results of the first order (Navier-Stokes)
hydrodynamics are obtained by neglecting the
$\dot{\pi}$ and $\dot{\Pi}$ terms in the above equations. Then one considers
$\pi^{(1)}=4\eta/(3\tau)$ and $\Pi^{(1)}=-\zeta/\tau$ and observes a re-heating
of the expanding quark gluon plasma:
\be
 \dot{e} = - \frac{4}{3\tau} e + \frac{4\eta}{3\tau^2} + \frac{\zeta}{\tau^2}.
\ee{EXPAND}
In particular for  early enough times the terms on the right hand side, being proportional
to $1/\tau^2$, dominate the evolution \cite{Muronga_old,MuroRi}.

For the radiative equation of state, $s=\frac{4}{3}\sigma^{1/4} \ep^{3/4}$, and
$e+p = 4\ep^2/(3e)$.
According to \re{TENSORSHEAR}, $\ep^2 = e^2 -3\Pi^2 - 3\pi^2/2$, thus $e+p = 4e/3 - 4(\Pi^2 + \pi^2/2)/e$.
Therefore, the cooling
due to the expansion is reduced, the dissipative terms physically re-heat the system.
There is nothing artificial about it.
This is fact more apparent for the proper time derivative of the effective internal energy density together with
eqs. (\ref{EVOLQGP}), where the terms $\Pi/\tau$ and $\pi/\tau$ do not appear explicitly,
\be
 \dot{\ep} = - \frac{4}{3\tau} \ep + \frac{e}{\ep}
    \left(\frac{3\pi^2}{4\eta} + \frac{\Pi^2}{\zeta}\right) \, .
\ee{TILDEERAD}
This form of the cooling equation reflects the fact that the source of physical re-heating consists of
the dissipative terms only, quadratic in dissipating momentum fluxes and inversely proportional
to the linear response coefficients.
For the sake of simplicity in the following we neglect the bulk viscosity, therefore eqs. (\ref{EVOLQGP}) reduce to
 \begin{eqnarray}
 \dot{\ep} &=& - \frac{4}{3\tau} \ep +
    \frac{3\pi^2}{4\eta}\sqrt{1+\frac{3}{2}\left(\frac{\pi}{\ep}\right)^2} \, ,\label{EqtoSolve1}\\
\dot{\pi} &=& \ep\left(\frac{2}{3\tau}-\frac{\pi}{2\eta}   \right)\sqrt{1+\frac{3}{2}\left(\frac{\pi}{\ep}\right)^2} \, .\label{EqtoSolve2}
\end{eqnarray}

These equations augmented with the radiation EOS are to be solved.
In the following, we will show the numerical results for the above equations in case of an ideal QGP with
3 massless quarks and 16 gluonic degrees of freedom, where the Stefan-Boltzmann coefficient is given as
$\sigma = 47.5\pi^2/30$, while the coefficient of viscosity is
$\eta = \eta_0 s = \eta_0 \frac{4}{3}(\sigma \ep^3)^{1/4}$.
Without viscous pressure terms, for and equilibrium EOS, one has to replace $\ep$ with $e$.
The initial conditions are given following Ref. \cite{SonHei07m}.
The starting time for the hydrodynamical evolution is, $\tau_0 = 0.6$ fm/c, the initial energy density
$e_0 = \ep_0 = 30$ GeV/fm$^3$, the initial viscosity to entropy density is $\eta_0 = 0.4$, while the
initial shear for  , $\pi_0 = 0$.
Here we note that other initial conditions and initial values are also possible, see for example \cite{Sat92a, Dumitru}.

To calculate the temperature eqs. \re{EqtoSolve1}-\re{EqtoSolve2} are solved together with the standard QGP EOS in four special cases. We give solutions for a perfect fluid, where $\pi \equiv 0$; for a first order dissipative fluid, i.e., the relativistic Navier-Stokes equations (NS),
where $\pi = 4\eta/3\tau $ and $ \ep=e$; for second order dissipative fluid, i.e., Israel-Stewart type fluid (IS),
where we set  $ \ep=e$ and four our higher order equations (HO) without any of the previous simplifications.    The initial shear for the transport equation is, $\pi_0 = 0$.
The corresponding thermodynamic temperatures are denoted by $T_{ID}$ for a perfect fluid, $T_{NS}$ for the NS equations,
$T_{IS}$ for the IS type equations and $T_{HO}$ and $\theta_{HO}$ for the two temperatures of our  higher order theory.
The existence of these temperatures is the outcome of the thermodynamical classification and treatment
as given in eqs. (\ref{no1}, \ref{DIFFENT}). Their different  physical role is clear from the structure of the theory;  $T$ is responsible for the equilibration and $\theta$  appears in the EOS. 

\begin{figure}[hbt!]
\centering
\includegraphics[width=12cm]{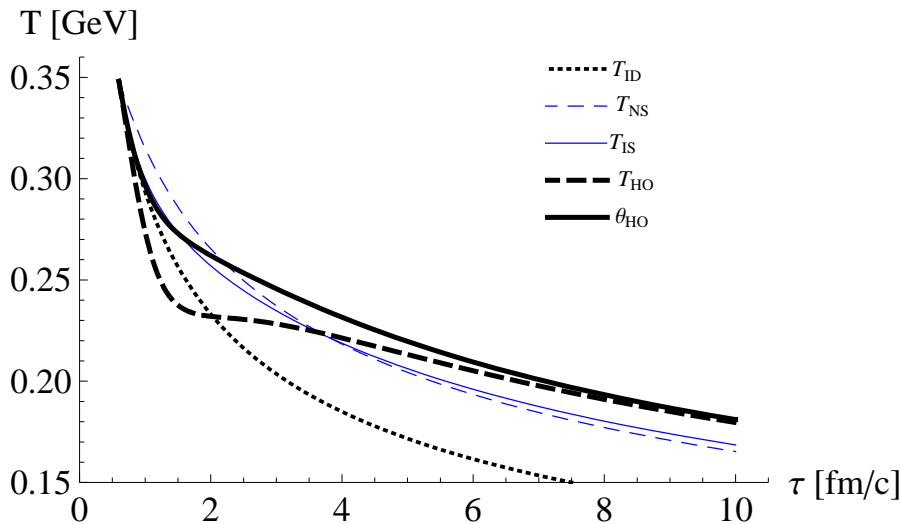}
\caption{The evolution of temperature as function of the proper time for the initial conditions and equations given in the text.
The dotted line shows the cooling of a perfect fluid, $T_{ID}$, the thin dashed and thin full lines are for
the NS and IS equations, denoted by $T_{NS}$ and $T_{IS}$.
The thick dashed and thick full lines are for the two temperatures of the HO equation, respectively denoted by $T_{HO}$ and $\theta_{HO}$.} 
\label{fig_1}
\end{figure}

On Fig. \ref{fig_1}, the evolution of the temperature is shown for the different simplified models. The overall behavior of the results is closely the same. For a longer time cooling is fastest for the ideal fluid, without dissipation, slower for the NS and IS fluids and slowest for the HO\ fluid, due to the reduced value of the expansion strength, $h = 4e/3 - 4(\Pi^2 + \pi^2/2)/e$,
compared to the case when, $h = 4e/3$.
The equilibrating temperature $T$ and the effective temperature $\theta$ tend to each other with the decreasing dissipation. 
For the HO solution, the equilibrating temperature decreases faster since $T= \theta \ep (\ep^2+1.5\pi^2)^{-1/2}$ and $\pi$ initially rises then decreases to the Navier-Stokes limit.

To phenomenologically study the  stability of the system, investigation related to the Reynolds number is standard practice.
The inverse Reynolds number, $R^{-1} = \pi/(e+p)$, is the ratio between dissipative and non-dissipative quantities, i.e.,
the ratio of dissipation to the strength of expansion.
The equation for the energy density in (\ref{EVOLQGP}) can be re-written in the following form,
\be
\dot{e} = \frac{h}{\tau} \left( 1 - R^{-1} \right) \, ,
\ee{reynolds_1}
where $h \equiv e + p$.
From eq. (\ref{TILDEERAD}) we get
\be
\dot{\ep} = \frac{h}{\tau} \left( 1 - R^{-1}_{\ep} \right) \, ,
\ee{reynolds_2}
where 
\be
R^{-1}_{\ep} = \frac{9\tau e \pi^2}{16 \eta \epsilon^2}  .
\ee{reynolds_3}
We easily realize that in both cases the energy increases as long as the inverse of the Reynolds number is smaller than one.
This leads to phenomenological upper bounds for the dissipative pressure, i.e., $\pi \leq 4\hat p$.

For the first order theory of Eckart one can show that in case the dissipative pressure becomes larger, initially or otherwise, than four times of the isotropic pressure, the solution to the equations becomes unstable \cite{Muronga_old,KouAta90a,Gyulassy_1}.
There reheating is closely related to stability. According to our knowledge the upper bound on dissipative quantities is not explicitly or quantitatively well specified for higher order theories,
therefore their domain of applicability and the stability conditions of the Bjorken flow and its generalizations are not clear.
  
\begin{figure}[hbt!]
\centering
\includegraphics[width=12cm]{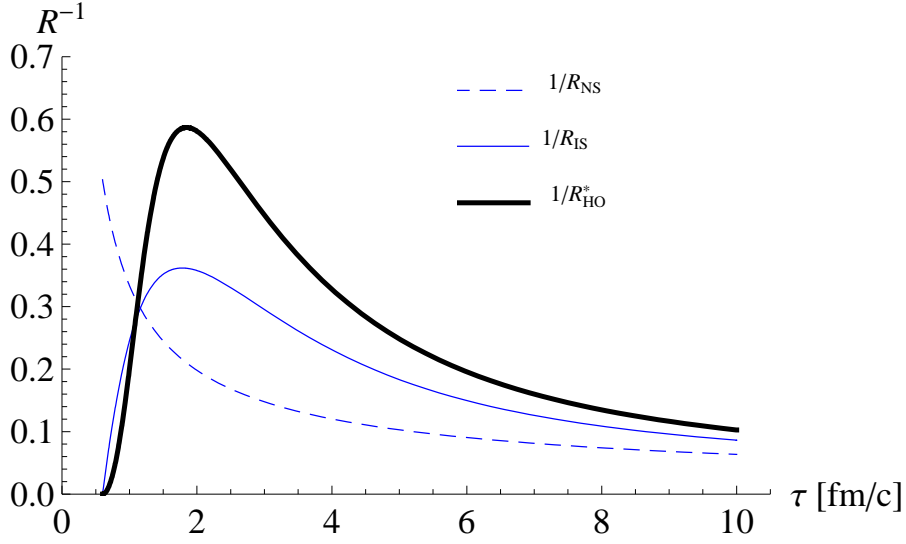}
\caption{The evolution of the inverse Reynolds number.
The thin dashed and thin full lines, denoted by $1/R_{NS}$ and $1/R_{IS}$, are for the NS and IS equations in case of the standard EOS.
The thick full line, denoted by $1/R_{HO}$ is for the modified Reynolds number \re{reynolds_3} in the higher order theory.
}
\label{fig_2}
\end{figure}

On Fig. \ref{fig_2}, the evolution of the inverse Reynolds number is shown for the two definitions given before.
The dashed line is for the NS equations while the full lines for the IS equations.
In the NS case the inverse Reynolds number decreases from its initial value and asymptotically approaches the perfect fluid limit due to 
the decrease of the expansion rate. 
For the IS and HO solutions, the system initially cannot compete with the fast expansion and first departs from equilibrium, 
and only later relaxes to the NS solutions.
Finally, we may conclude that there is no reheating in any of above presented cases because the inverse of the Reynolds number
is less than one.

Because our modified Reynolds number is explicitly time dependent, the separation of cooling and reheating solutions is not straightforward without the solutions of the equations.  In fig. \ref{fig_3} we compare the reheating capabilities of the different theories. The minimal initial energy density for a reheating solution is plotted as function of starting proper time in case of zero initial viscous shear pressure. The shear viscosity was $\eta_0=0.3$ in the left figure and $\eta_0=0.08$ at right. For the first order theory of Eckart one can give the explicit condition as $e_0 = \sigma \left(\frac{4\eta_0}{3\tau_0}\right)^4 $. For the Israel-Stewart and four our higher order theory the corresponding curves can be well approximated as $e_0 =b\tau_0^{-4}$, where the $b$ parameter values are tabulated  in table 1.

{\center \begin{tabular}{c||c|c|c}
  $\eta_0$ & Eckart & Israel-Stewart & This paper  \\\hline\hline

0.3  & $6\cdot 10^{-4}$ & $5.6\cdot 10^{-7}$ & $2.67\cdot 10^{-4}$ \\\hline
0.08 & $3\cdot 10^{-6}$ & $2.89\cdot 10^{-9}$ & $1.75\cdot 10^{-4}$ \\
\end{tabular}\label{tab_1}

{TAB. 1: $b$ {\small [$GeV fm/c^4$]} parameter values characterizing the conditions of reheating for the different dissipative fluids.}}
\begin{figure}[hbt!]
\centering
\includegraphics[width=8cm]{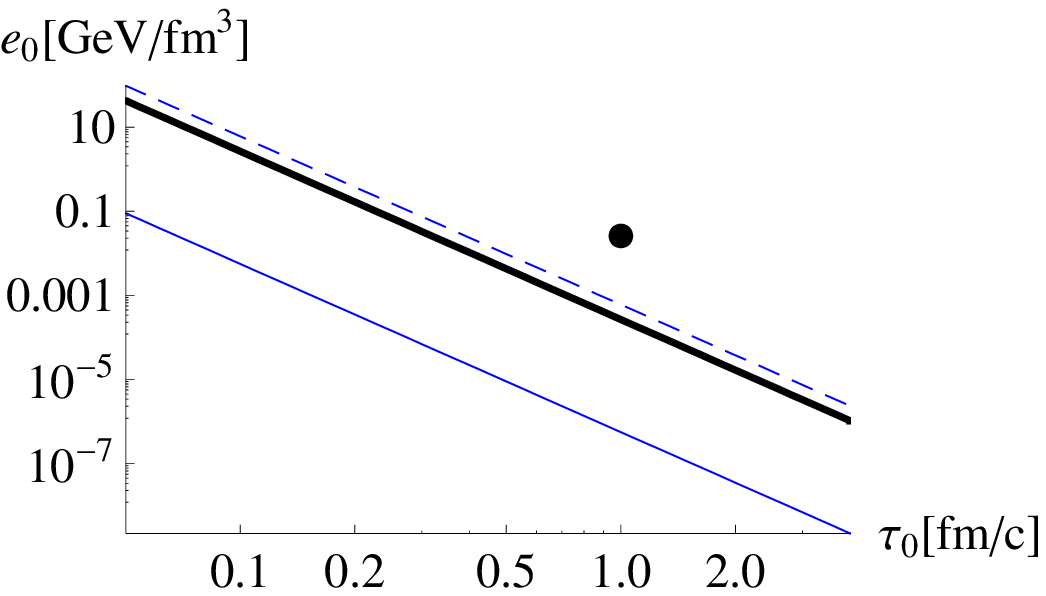}
\includegraphics[width=8cm]{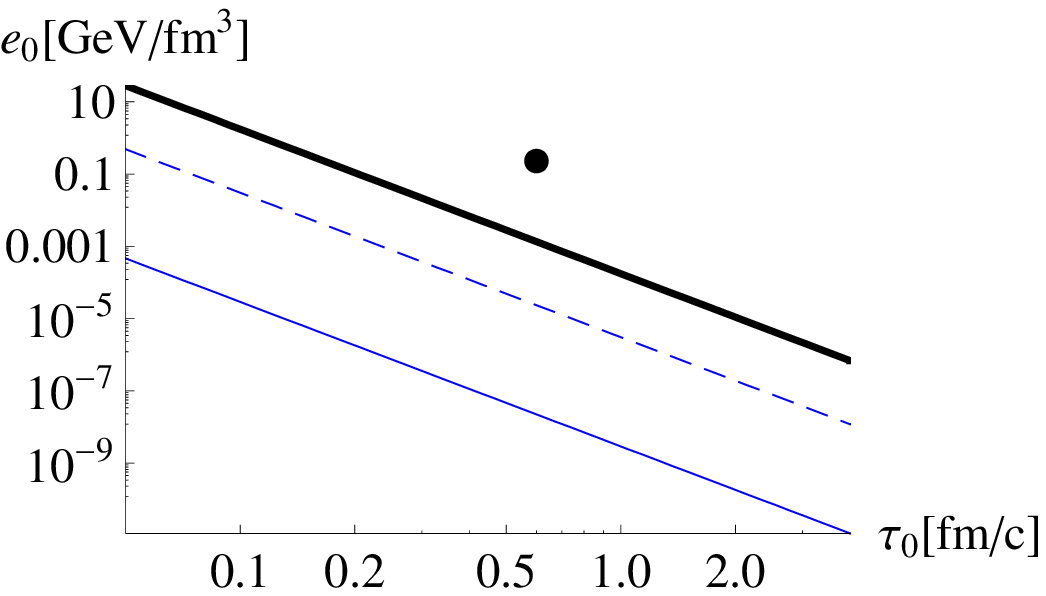}
\caption{The initial energy density to get a reheating flow as a function of the starting proper time for $\eta_0=0.3$ to
the left and $\eta_0=0.08$ to the right.
The thin dashed and thin full lines are for the NS and IS fluids and the thick full line is for the HO model.
Initial conditions below the lines lead to  reheating solutions. The dots indicate initial conditions considered to be realistic for the RHIC and LHC experiments \cite{SonHei07m,Sat92a}.}
\label{fig_3}
\end{figure}

\section{Discussion}

In our approach we have fixed the thermodynamic coefficients, akin to the \(\alpha_i\) $\beta_i$ coefficients of the
Israel-Stewart theory, respectively or the \(\alpha\) coefficient of the \"Ottinger-Grmela theory \cite{Ott05b,DusTea07m}
to particular values, that otherwise could have been calculated from kinetic models.
There are several reasons why a direct phenomenological approach can be fruitful in heavy ion physics.

\begin{enumerate}
 \item The available transport coefficients calculated from kinetic theory are not allways realistic and sometimes are controversial.
Not realistic, when they are related to oversimplified microscopic properties (like one component ideal gases) \cite{Israel_Stewart}.
In this approach some structural properties are considered for the second order coefficients \cite{Muronga_new}.

 \item Certain results of hydrodynamic calculations are independent of the exact values of the second order coefficients: A difference in the initial values of energy density, equilibrium and dissipative pressures can be compensated by an appropriate choice of the initial proper time ($\tau_0$) in the Bjorken flow scenario. We have argued that first order theories cannot be excluded by causality reasons, because the actual violation of causality may
be beyond the validity range of the hydrodynamic approach
\cite{BiroVan}. Baier et. al. has been arguing that the difference of the solutions of a first order and a second order theory do appear at microscopic distances, beyond the validity range of hydrodynamics \cite{BaiAta07m}. Several   theories of relativistic dissipative fluids (e.g. Israel-Stewart, the divergence type Geroch or M\"uller-Ruggieri theories, Baier et al., Koide et al., \"Ottinger-Grmela) may belong to this class.

 However, it is easy to see, that generic instabilities of the corresponding theory could destroy the above argumentation. Up to know the instability of the first order theories and stability conditions in Israel-Stewart
theory \cite{HisLin83a,HisLin87a} has been well investigated and for the $\pi=\Pi = 0, q=0$ minimal version of our higher order approach were pointed out \cite{BiroVan}. According to our knowledge, the conditions of stability of homogeneous solutions are not known for the rest of the above mentioned theories.

\item There is no a priori reason why one should end at second order
extension of the first order theory, beyond convenience.
\end{enumerate}

Within our thermodynamic framework, not showing generic instabilities for $q^i\neq 0$ we have investigated the role of heat conduction together with a generalized, longitudinally accelerating Bjorken flow in more detail. Our conclusion was that for stationary flow the heat flux must vanish also in a general frame (also in the Eckart frame), there is no need for the the customary Landau-Lifshitz condition ($q^i=0$). On the other hand the reheating effect is a consequence of dissipative physical phenomena, not of generic instabilities.

A peculiar property of our approach to the relativistic internal energy is the distinctness of the derivatives of the entropy with respect the total or internal energies. This fact resulted in a doubled set of non-equilibrium intensive thermodynamic quantities according to the respective derivatives ($ T$ and $\theta$, $\mu$ and $\hat\mu$, $p$ and $\hat p$). We have seen that the intensive variables related to the internal energy (\(\theta,\hat\mu, \hat p \)) resulted in more natural thermodynamic relations:\ The intensive variables related to the total energy ($T,\mu,p$ and the coefficients of the viscous pressure terms in the Gibbs relation \re{DIFFENT}) are rather related to the  energy equilibration due to dissipative processes. In particular the gradient of $T$ appeared in the generalized Fourier law \re{Four} but for the radiation thermodynamics it was most straightforward to consider $\theta$ as temperature.

\section{Acknowledgement}

This work has been supported by the Hungarian Scientific Research fund (OTKA K49466, T48489),
by a common project of OTKA and the National Office for Research and Technology (OTKA/NKTH 68108)
and by a DFG-MTA program.
TSB thanks for the hospitality of the Yukawa Institute,
PV acknowledges the support of a Bolyai scholarship of the Hungarian Academy of Sciences,
and EM thanks for the Alexander von Humboldt foundation for support.


\end{document}